# Viewpoint: On the Emergence of van der Waals Magnets: A Personal Reflection

Je-Geun Park

Department of Physics and Astronomy, Seoul National University, Seoul 08826, Korea

* Email: jgpark10@snu.ac.kr

**Abstract**: The observation of magnetism in atomically thin van der Waals (vdW) antiferromagnets ($FePS_3$, $NiPS_3$, and $MnPS_3$) in 2016 marked an important moment in the development of two-dimensional (2D) physics. In this personal reflection, I describe how a simple question, posed in the early 2010s, motivated experimental efforts that culminated in the demonstration of antiferromagnetic order in monolayer $FePS_3$. Alongside subsequent reports of vdW ferromagnets in 2017, these developments helped establish intrinsic magnetism as a viable degree of freedom in atomically thin materials. I close with personal lessons drawn from this period and a perspective on the opportunities that now shape the field's second decade and beyond.

At the beginning of every scientific journey lies a question. The best ones appear almost disarmingly simple, yet their consequences can ripple across entire fields. The question was a simple one: Can a monolayer magnet sustain an (anti)ferromagnetic ground state? In hindsight, it may appear obvious in light of seminal theoretical work by many leading scientists since the 1940s. However, in reality, it required years of detours, doubts, and surprises. That single question became the thread that wove together much of my work over the past 15 years.

Part of the beauty of such a question lies in its simplicity and universality. Many physicists must have pondered the fate of magnetic order in the strict two-dimensional limit before me, especially in the wake of Onsager's solution of the Ising model and the Mermin–Wagner theorem. Yet for decades, this remained a largely theoretical curiosity. The materials platforms available to

us—bulk magnets or quasi-2D compounds—were never quite suited to test the idea directly. It was only with the rise of graphene and the larger family of exfoliable van der Waals crystals that the question could be posed again, this time with the tools to seek an experimental answer.

When I discussed this problem publicly at the Korean Physical Society meeting in 2015 [1], it felt both modest and audacious at the same time. Modest because the phrasing was simple; audacious because the implications were profound. If magnetism could be stabilised in a true vdW monolayer, it would open an entirely new arena for two-dimensional physics, with consequences stretching from fundamental theory to spintronics and quantum devices. At that moment, I could not have foreseen how quickly events would unfold, nor how this deceptively straightforward question would reshape the direction of my scientific path.

For many of us, the Mermin–Wagner theorem had come to symbolise the fragility of long-range order in two dimensions [2]. Yet, the rise of 2D materials after graphene made the question impossible to ignore. What seemed, at first, an unfashionable idea soon became an obsession: identifying a credible material platform, isolating the monolayer, and testing whether intrinsic magnetic order could be stabilised. With hindsight, those early talks, in the KPS meeting in 2015 and at a meeting at the University of Tokyo [3] a few months later, marked a shift in my own thinking. Colleagues who attended either of my talks expressed interest, but needless to say, they were all cautious, on the edge of scepticism; they all asked for the clear-cut proof. This challenge clarified the task and sharpened the course of the following work.

The choice of platform, in this case, was crucial for the materials used. Transition-metal phosphorus trisulfides, $TMPS_3$ (TM = Mn, Fe, Co, Ni), provided a natural playground: layered structures and readily exfoliable materials. Most importantly to me, they realise all three distinct spin Hamiltonians: Ising, XY and Heisenberg models, only with varying transition metal atoms [4]. My group began serious efforts around 2013 and by 2016 had demonstrated that antiferromagnetic order in $FePS_3$ persists in the monolayer limit [5]. These results are consistent with Onsager's solution of the two-dimensional Ising model [6]: robust 2D magnetism can be realised with properly symmetry-allowed anisotropies.

Around the same time, I wrote a JPCM Viewpoint [7] ("Opportunities and challenges of 2D magnetic van der Waals materials: magnetic graphene?"), capturing a moment when the field was pivoting from speculation about possibilities to demonstrations in hand. What followed confirmed that the phenomenon was not confined to a single compound or ordering pattern.

In 2017, reports emerged of monolayer ferromagnetism in vdW crystals such as $Cr_2Ge_2Te_6$ [8] and $CrI_3$ [9], establishing a broader class of 2D magnetic matter, alongside work on $FePS_3$. Soon after, metallic vdW magnets such as $Fe_3GeTe_2$ were shown to host tunable ferromagnetism down to the few-layer limit, with electric-field control of magnetic anisotropy [10]. Beyond those early materials, the landscape expanded to include topological magnets ($MnBi_2Te_4$ and $Co_{1/3}$-$TaS_2$) [11-13], multiferroics ($NiI_2$) [14, 15], and air-stable monolayer ferromagnets (CrSBr and $CrPS_4$) [16, 17]. Even more recently, moire engineering in twisted bilayers has revealed emergent skyrmionic textures and noncollinear multi-Q states [18, 19].

Global efforts expanded the frontiers of vdW magnetism with several notable advances: the identification of new families of 2D (anti)ferromagnetic materials; mapping of magnetic excitations and anisotropies; and vdW heterostructures with semiconductors, superconductors, and topological materials. Through these developments, the field matured rapidly from the demonstration of existence to the engineering of coupling, proximity effects, and control, all within a few years.

Looking back, the first decade, from 2016 to 2025, established the existence of vdW magnets. I

anticipate that the next decade will focus on designing ever more novel materials systems, both artificial and natural. For example, it will be exciting to build heterostructures where magnetism couples to superconductivity or to topological edge states, and to explore proximity effects in graphene or TMDs gated by vdW magnets. Another especially promising direction is to interface vdW magnets with more conventional magnetic platforms, notably oxides [20]. I predict that it will be extremely fertile ground, as oxide magnetism is one of the most commonly available forms of magnet and one of the most studied materials over the years, particularly in the last four decades. The challenge now is not whether 2D magnetism exists, but how to control it with precision and integrate it into quantum and spintronic devices.

Looking back, three lessons stand out.
- Unfashionable seeds can grow. The initial talks in 2015 & 2016 were modest, and the materials were, at the time, peripheral to mainstream 2D research. Yet those were exactly the conditions under which a new research direction could take root.
- Persistence matters. Years of steady work on $TMPX_3$ compounds paid off precisely because the platform was ready when the right question was asked.
- Scepticism is critical. Colleagues' constructive doubt led to clearer experiments and better-focused narratives.

Looking back, the first decade of vdW magnetism was characterised by increasingly compelling demonstrations. Several groups, including my own, showed that long-range order can be stabilised in vdW monolayers of both antiferromagnetic and ferromagnetic materials. And the materials base is broad enough to sustain an entire research landscape.

With those progresses over the last decades, the central question has since shifted significantly [21].

No longer is the question "Does 2D magnetism exist?" but "How can it be controlled, coupled, and harnessed?" Twisted and lattice-matched heterostructures, chiral and nematic orders, noncollinear multi-Q textures, and topological responses now define many new frontiers of van der Waals magnetism. Future progress will depend on quantitative control of anisotropy, interlayer exchange, moire periodicity, and interfacial symmetry alongside advances in materials growth and nanoscale characterisation. Therefore, if the first decade established existence, the next will be about designing new 2D magnetic systems: writing spin into the vdW toolkit with the same finesse that charge and valley have enjoyed.

I thank the past and present members of my group and collaborators who shared my passion for a simple yet crazy idea, as well as colleagues whose questions sharpened my personal quest over the years. The work was supported by the Leading Researcher Program of the National Research Foundation of Korea (Grant No. RS-2020-NR049405).